# Correlated Node Behavior Model based on Semi Markov Process for MANETS

A.H Azni[1], Rabiah Ahmad[2], Zul Azri Muhamad Noh[3], Abd Samad Hasan Basari[4], Burairah Hussin[5]

[1,2,3,4,5]Universiti Teknikal Malaysia Melaka
Faculty of Information and Communication Technology,
76100 Durian Tunggal, Melaka, Malaysia

**Abstract**
This paper introduces a new model for node behavior namely Correlated Node Behavior Model which is an extension of Node Behavior Model. The model adopts semi Markov process in continuous time which clusters the node that has correlation. The key parameter of the process is determined by five probabilistic parameters based on the Markovian model. Computed from the transition probabilities of the semi-Markov process, the node correlation impact on network survivability and resilience can be measure quantitatively. From the result, the quantitative analysis of correlated node behavior on the survivability is obtained through mathematical description, and the effectiveness and rationality of the proposed model are verified through numerical analysis. The analytical results show that the effect from correlated failure nodes on network survivability is much severer than other misbehaviors.
**Keywords:** network resilience, survivability, node behavior model, traffic profiling, MANETS.

## 1. Introduction

The node behavior pattern plays an important role in performance analysis of mobile and wireless networks. In mobile networks, for example, a node may change its behavior from normal to misbehave node which directly affects the connectivity and availability of the network [1]. The impact of misbehave node is quite challenging due to multiple failures caused by node mobility, energy depletion and Denial of Services (DoS) attacks. In addition, node behavior has major effect on route discovery, packets forwarding, and network control message [2][3][4]. However, nodes in the network not only affect individual node, but it may affect multiple nodes which has a direct or indirect correlation between nodes.

The modeling of correlated node behavior is thus an essential part in analyzing and designing survivability framework in wireless network. In this paper, Markov process is represented to model and analyze the stochastic properties of the node's behavior. We also design a correlated node behavior model that could dynamically affect network resilience in MANETS. The rest of the paper is organized as follows Section 2 gives related works of node behavior theory. Section 3 defines node behavior in MANETS. Sections 4 presents a proposed model defining four-sate transition of correlated node behaviors and describe probabilistic theory to model correlated nodes behavior. Section 5 discusses the parameters use as a performance evaluation. Finally we conclude our paper.

## 2. Related Works

The studies of node behavior have helped many researchers to understand many research problems such as the design of fault tolerant routing protocol and analysis of network performance, thoroughly. Currently, node behavior modeled by [5][6] profile the normal behavior of wireless network. It used genetic algorithm to learn current behavior based on past history. This approach has been used widely in intrusion detection system to detect anomalies in the network. These model measures network performance based on single or individual node. As mention by [7], individual node may not accurately measure real life application. It needs to take into account an impact of correlated node instead of individual node. Work in [8][9][10] deals with modeling of selfish node behavior using game theory. Game theory is a branch of economics that deals with strategic and rational behavior [11]. It appears to be a natural tool for both designing and analyzing the interaction among players/nodes. The theory seems suitable to model selfish behavior where energy conservation is an issue to selfish node. It tends to limit their support to other nodes as this will costs energy. Thus, to stimulate cooperation, the nodes need to be rewarded every time it cooperates. The other theory used to model behavior is Markov chain theory [12][13][14]. Markov chain is a mathematical system that undergoes transition from one state to another in a chain manner [15]. Markov chain theory suits for nodes in ad hoc network as it is general enough to capture the major characteristics and yet simple enough to formulate nodes long-run behavior [16][17]. To the authors knowledge, neither one of the research in this area involve correlated nodes behavior in their studies.

Work done in this paper applied Semi Markov to determine node behavior transition. Semi Markov is a probabilistic system that made its transitions according to transition probability matrix of Markov process, but whose time between transitions could be an arbitrary random variable which depends on the transition. This work is an extension of a research done by [7] which model network survivability based on individual node behavior. He assumed that node behavior has no correlation with another node's behavior in wireless network. Thus, his model is tractable, however, this assumption may not always hold in real environment. In real environment node has a correlation with other nodes in such a way that if a node has more and more neighbors failed, it may need to load more traffic originally forwarded by those failed neighbors, and thus might become failed faster due to excessive energy consumption. Similarly, it is also possible that the more malicious neighbors a node has, the more likely the node will be compromised by its malicious neighbors. Therefore, the paper examines how these correlated behaviors will affect the network resilience.

## 3. Node Behavior Definitions

In MANETS, the nodes are dynamically and arbitrarily change its behavior from cooperative to misbehave node. Based on [9] node behaviors in MANETS are classified into four types of behaviors. They can be classified as:

- *Cooperative Nodes* are active in route discovery and packet forwarding, but not in launching attacks
- *Failed Nodes* are not active in route discovery
- *Malicious Nodes* are active both in route discovery and launching attacks
- *Selfish Nodes* are active in route discovery, but not in packet forwarding. They tend to drop data packets of others to save their energy so that they could transmit more of their own packets and also to reduce the latency of their packets.

Whenever node joints the network, it is assume as normal or cooperative. It may change its behavior to misbehave node due to various reasons. In this model, we specify rules for a node transition. Figure 1 shows the transition of node behavior in MANETS.

a) Cooperative node (C) may change its state either to selfish, malicious or failure node. At cooperative state, it is exposed to become failed due to energy exhaustion, misconfiguration, and so on.
b) It is also possible to convert a selfish (*S*) node to be cooperative again by means of proper configurations. A selfish or cooperative node can become malicious due to being compromised or failed due to power depletion.
c) A malicious (*M*) node can become a failed node (*F)*, but it will not be considered to be cooperative or selfish any more even if its disruptive behaviors are intermittent only.
d) A failed node (*F)* can become cooperative again if it is recovered and responds to routing operations.

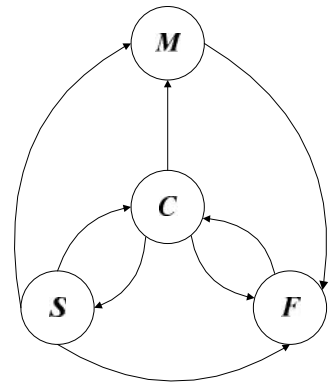

Fig. 1: Node behavior transition

## 4. Correlated Node Behavior Model

In this section, we use a Semi-Markov process to model node behavior transitions, analyzed the stochastic properties of node behavior, formulating the transition matrix and model state transition for correlated node behavior.

4.1 Semi Markov and Stochastic Properties Node Behavior Model

Due to the fact that node in MANETS is more inclined to be failed over time, we find that, the probability changes its behavior dependent on time. Therefore, node transition cannot simply described by Markov chain because of its time-dependent property. The semi-Markov process denoted by:

$$Z(t) = X_n, \quad t_n \leq t < t_{n+1} \quad (1)$$

with a state space S = *{C (cooperative), S (selfish), M (malicious), F (failed)}*. $X_n$ denotes the *embedded* Markov chain of Z(*t*), which has a finite state space S**,** and the *n*th state visited [18].Thus, $X_n$ is *irreducible* and *ergodic*. By

Collolary 9 -11 (pp 325) in [7] we know that $Z(t)$ is irreducible and $Z(t)$ is the state of process at its most recent transition. The transition probability from state $i$ to state $j$ is defined as follows:

$$P_{ij} = \lim_{t \to \infty} \Pr(X_{n+1} = j, t_{n+1} - t_n \leq t \mid X_n = i)$$
$$= \Pr(X_{n+1} = j \mid X_n = i)$$

(2)

Then a matrix $\overline{P} = (P_{ij})$ is the transition probability matrix (TPM) of $\{X_n\}$. The construction of $\overline{P}$ can be determined by the observation of empirical results. The state transition diagram of semi Markov node behavior model is shown in Fig.2, which is determined by characteristics of node behaviors.

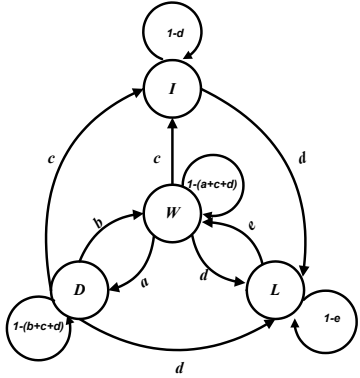

Fig. 2: Semi-Markov Process for Node Behavior

A node in proposed model is viewed as having in four states namely forward (W), drop (D), inject (I) and loss (L) which adequately describe the behavior of the node during forwarding, dropping, injecting and loss state. Let the probability of dropping packets due to selfishness and the probability of forwarding packets due to altruistic nature shown by the node $a$ and $b$ respectively and they are independent of each other. Let $c$ be the probability of injecting packets due to malicious activity and $d$ is the probability of loss packets due to exhausted battery power, out of transmission range or malfunction. Probability of recovery shown by $e$ define the recovery of node from failure state to cooperative again after it has been recovered, recharged or repaired. To determine the current behavioral status of neighbor node $m$ at instant time $t$ is formally given as:

$$N^{(m)}(t) \begin{cases} W, \text{ if the level of dropping, a} < \text{threshold and level of forwarding b > threshold; and e=1} \\ D, \text{ if level of dropping a > threshold;} \\ I, \text{ if the injecting level, c > threshold and level of forwarding b<threshold} \\ L, \text{ if level of loss d=1;} \end{cases}$$

(3)

### 4.2 Formulation of Correlated Node Behavior Transition Probability Matrix (TPM).

Based on node behavior above, transition probability matrix (TPM) of $\{X_n\}$ is given below:

$$P = \begin{array}{c} \\ W \\ D \\ I \\ L \end{array} \begin{array}{cccc} W & D & I & L \end{array} \\ \begin{bmatrix} 1-(a+c+d) & a & c & d \\ b & 1-(b+c+d) & c & d \\ 0 & 0 & 1-d & d \\ e & 0 & 0 & 1-e \end{bmatrix}$$

(4)

The "0" in the matrix means that it is not possible to make transition between the two states based on the rules specify in section 3.0. Since it is a stochastic matrix, the summation of transition probabilities to a state must be equal to 1. $\{Z(t)\}$ is also associated with the time distributions between two successive transitions. Let $T_{ij}$ denote the time spent in state $i$ given the next state $j$. Then $Z_{ij}(t)$ is a commonly used notation for cumulative distribution function (CDF) of $T_{ij}$, defined by :

$$Z_{ij}(t) = \Pr(T_{ij} \leq t)$$
$$= \Pr(t_{n+1} - t_n \leq \mid X_n = i, X_{n+1} = j)$$

(5)

where $i,j \in S$.

*Proof: Let $P_{ij}$ be the parameter for a, b, c d, and e define above for the node $N^m$ at time instant t, we can specify:*

$a = P[N^{(m)}_{(t-1)} = C \mid N^{(m)}_{(t)} = S]$
$b = P[N^{(m)}_{(t-1)} = S \mid N^{(m)}_{(t)} = C]$
$c = P[N^{(m)}_{(t-1)} = C \mid N^{(m)}_{(t)} = M]$ *or*

$P[N^{(m)}_{(t-1)} = S \mid N^{(m)}_{(t)} = M]$
$d = P[N^{(m)}_{(t-1)} = C \mid N^{(m)}_{(t)} = F]$ or $P[N^{(m)}_{(t-1)} = S \mid N^{(m)}_{(t)} = F]$
or $P[N^{(m)}_{(t-1)} = M \mid N^{(m)}_{(t)} = F]$
$e = P[N^{(m)}_{(t-1)} = F \mid N^{(m)}_{(t)} = C]$

(6)

By knowing the transition probability $P_{ij}$ and transition time distribution $Z_{ij}(t)$, which are defined above, the transient distribution $P_{ij}$ converge to a limiting $P_i$ can be calculated by

$$P_i = \lim_{t \to \infty, \forall j \in S} \Pr(Z(t) = i \mid Z(0) = j) = \frac{\pi_i E[T_i]}{\sum_j \pi_j E[T_j]}$$

(7)

where $\pi_i$ is the stationary probability of state $i$ of $X_n$. To obtain the steady state probabilities, we need to solve the equation

$$\Pi \cdot P = \Pi$$

$\sum_{i \in S} \pi_i = 1,\ \pi_i \geq 0$, and $E[T_i] = \sum_{j \in S} P_{ij} E[T_{ij}]$

(8)

Let $\Pi$ is the probability vector in steady state and **P** is the matrix representing the transition probability distribution from Fig.3. Thus, to calculate $P_i$, we only need to estimate $P_{ij}$ and $E[T_i]$, which are normally easier to obtain from statistic, then we can use equation in (8) to obtain $\pi_i$ and $E[T_i]$. After $\pi_i$ and $E[T_i]$ are obtained, we can use equation (7) to derive $P_i$.

*Proof: First, for the given embedded Markov Chain of $\{X_n\}$ associated with the state space S and TPM **P** defined by (4), it is trivial to prove that $\{X_n\}$ is irreducible and positive recurrent. Thus, the SMP $\{Z(t)\}$ is irreducible. Second, based on our assumptions in Section 3.0, a node works in any behavior state for a finite time, which implies $E[T_i] < \infty$ and $\sum_{i \in S} E[T_i] < \infty$. Thus, the SMP $\{Z(t)\}$ is positive recurrent as well. Finally, by Theorem 9-3 [19] the limiting probability exists and can be given by (7).*

In order to formulating TPM for correlated node behavior, let $N^1$ and $N^2$ are two nodes connected in a network. The corresponding TPM for $N^1$ and $N^2$ are given below:

Table 1a: Node $N^1$

|   | W | D | I | L |
|---|---|---|---|---|
| W | 1-(a+c+d)$^1$ | a$^1$ | c$^1$ | d$^1$ |
| D | b$^1$ | 1-(b+c+d)$^1$ | c$^1$ | d$^1$ |
| I | 0 | 0 | 1-d$^1$ | d$^1$ |
| L | e$^1$ | 0 | 0 | 1-e$^1$ |

Table 1b: Node $N^2$

|   | W | D | I | L |
|---|---|---|---|---|
| W | 1-(a+c+d)$^2$ | a$^2$ | c$^2$ | d$^2$ |
| D | b$^2$ | 1-(b+c+d)$^2$ | c$^2$ | d$^2$ |
| I | 0 | 0 | 1-d$^2$ | d$^2$ |
| L | e$^2$ | 0 | 0 | 1-e$^2$ |

With two nodes, the state space occupies sixteen finite states such as { WW, WD, WI, WL, DW, DD, DI, DL, IW, ID, II, IL, LW, LD, LI, LL} at any point of time. It requires $O(4^m)$ computation complexity, where $m$ refer to number of nodes in the network. However, for correlated nodes behavior, we proposed a clustering method which groups the nodes according its current status and behavior and thus, reduce the computational complexity. Fig 3 illustrated the clustering of correlated transition probability distribution between different correlated state and correlated PTM matrix is also given in table 2 below. The four correlated states behavior are grouped by $\{S_0, S_1, S_2, S_3\}$ mapped against $\{WW, DD, II, LL\}$.

Let $\{u, v, w, x\}$ be a function mapping states into correlated state behaviors. Denote $\pi_i$ the probability of being in steady state. If the state space is finite, then the equation in (8) can be solve to obtain $\pi_i$. The status of a node at current $t$ time is given by (3). Thus, the cluster of correlated node behavior involving two nodes $N^1$ and $N^2$ are:

$$N^{(1,2)}_{(t)} \begin{cases} S_0 = C, & \text{if } [(N^{(1)}_{(t)} = W) \text{ and } (N^{(2)}_{(t)} = W)] \\ S_1 = S, & \text{if } [(N^{(1)}_{(t)} = D) \text{ and } (N^{(2)}_{(t)}) = D)] \\ S_2 = M, & \text{if } [(N^{(1)}_{(t)} = I) \text{ and } (N^{(2)}_{(t)} = I)] \\ S_3 = F, & \text{if } [(N^{(1)}_{(t)} = L) \text{ and } (N^{(2)}_{(t)} = L) \end{cases}$$

(9)

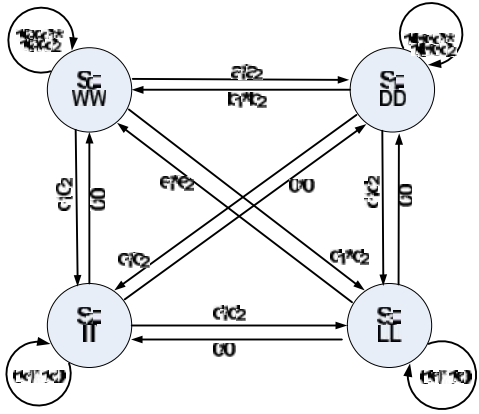

Fig.3: Correlated State Transition representing the node behavior

Table 2: Correlated PTM for four state Markov chain

|  | $S_0$ | $S_1$ | $S_2$ | $S_3$ |
|---|---|---|---|---|
| $S_0$ | $1-(a+c+d)^1$ * $1-(a+c+d)^2$ | $a^1*a^2$ | $c^1*c^2$ | $d^1*d^2$ |
| $S_1$ | $b^1*b^2$ | $1-(b+c+d)^1$ * $1-b+c+d)^2$ | $c^1*c^2$ | $d^1*d^2$ |
| $S_2$ | $0*0$ | $0*0$ | $(1-d)^1*(1-d)^2$ | $d^1*d^2$ |
| $S_3$ | $0*0$ | $0*0$ | $0*0$ | $(1-e)^1*(1-e)^2$ |

Table 2 is constructed to map PTM against correlated state transition $S_0$, $S_1$, $S_2$, and $S_3$ in Fig.3. We propose a correlated function based on conditional probabilities:

$$u = P[N^{(1...m)}_{(t-1)} = C \mid N^{(1...m)}_{(t)} = S]$$

$$v = P[N^{(1...m)}_{(t-1)} = S \mid N^{(1...m)}_{(t)} = C] \text{ or}$$

$$P[N^{(1...m)}_{(t-1)} = F \mid N^{(1...m)}_{(t)} = C]$$

$$w = P[N^{(1...m)}_{(t-1)} = C \mid N^{(1...m)}_{(t)} = M] \text{ or}$$

$$P[N^{(1...m)}_{(t-1)} = S \mid N^{(1...m)}_{(t)} = M]$$

$$x = P[N^{(1...m)}_{(t-1)} = C \mid N^{(1...m)}_{(t)} = F] \text{ or}$$

$$P[N^{(1...m)}_{(t-1)} = S \mid N^{(1...m)}_{(t)} = F] \text{ or}$$

$$P[N^{(1...m)}_{(t-1)} = M \mid N^{(1...m)}_{(t)} = F]$$

(10)

To show how correlated function works, let $N^1$ and $N^2$ be the connected node in the network. Resultant from equation (6) and (10), we get:

$$u^1 = P[N^{(m)}_{(t-1)} = C \mid N^{(m)}_{(t)} = S]$$

$$v^1 = P[N^{(m)}_{(t-1)} = S \mid N^{(m)}_{(t)} = C] \text{ or}$$

$$P[N^{(m)}_{(t-1)} = F \mid N^{(m)}_{(t)} = C]$$

$$w^1 = P[N^{(m)}_{(t-1)} = C \mid N^{(m)}_{(t)} = M] \text{ or}$$

$$P[N^{(m)}_{(t-1)} = S \mid N^{(m)}_{(t)} = M]$$

$$x^1 = P[N^{(m)}_{(t-1)} = C \mid N^{(m)}_{(t)} = F] \text{ or}$$

$$P[N^{(m)}_{(t-1)} = S \mid N^{(m)}_{(t)} = F] \text{ or } P[N^{(m)}_{(t-1)} = M \mid N^{(m)}_{(t)} = F]$$

Then, using equation (8), we can obtain $u^1$ as:

$$u^{(1)} = \frac{P[N^{(1)}_{(t-1)} = C \mid N^{(1)}_{(t)} = S]}{P[N^{(1)}_{(t-1)} = C]} * P[N^{(1)}_{(t)} = S]$$

$$= \frac{a^1}{1-(\pi_1 + \pi_2 + \pi_3)} * \pi_1$$

where $\pi_1, \pi_2, \pi_3$ are obtained from solving equation (8) for correlated state model shown in Fig.3. Rest of the function can be calculated in similar ways. The equation only represents a single pair of neighbor nodes only. As part of correlated state transition, its functions ($u^{(1,2)}$, $v^{(1,2)}$, $w^{(1,2)}$, $x^{(1,2)}$), can be represented as per equation (10). The next step, it is proposed to have an iterative approach to model the correlated nodes to logically present nodes as correlated clusters. Let $m \geq 2$ be the total number of neighbor nodes. In the first iteration, neighbor nodes $N^1$ and $N^2$ into equivalent node. Then the resulting equivalent node is combined with $N^3$, and so on, until all $m$ neighbor nodes are combined together to form a function of correlated node. This iteration is repeated till all nodes are clustered according to their correlated state. Using the resultant result from correlated state model in Fig.3 and equation set (9), the correlated function can be rewritten can as:

$$u^{(1...m)} = \frac{P\left[N^{(1...m)}_{(t-1)} = C \mid N^{(1...m)}_{(t)} = S\right]}{P\left[N^{(1...m)}_{(t-1)} = C\right]} * P\left[N^{(1...m)}_{(t)} = S\right]$$

$$= \frac{a^{1...m} * a^{1...m}}{1 - (\pi_1 + \pi_2 + \pi_3)} * \pi_1$$

$$v^{(1...m)} = \frac{P\left[N^{(1...m)}_{(t-1)} = S \mid N^{(1...m)}_{(t)} = C\right]}{P\left[N^{(1...m)}_{(t-1)} = S\right]} * P\left[N^{(1...m)}_{(t)} = C\right] +$$

$$\frac{P\left[N^{(1...m)}_{(t-1)} = F \mid N^{(1...m)}_{(t)} = C\right]}{P\left[N^{(1...m)}_{(t-1)} = F\right]} * P\left[N^{(1...m)}_{(t)} = C\right]$$

$$= \frac{b^{1.m} * b^{1.m}}{1 - (\pi_0 + \pi_2 + \pi_3)} * \pi_0 + \frac{e^{1.m} * e^{1.m}}{1 - (\pi_0 + \pi_1 + \pi_2)} * \pi_0$$

$$w^{(1...m)} = \frac{P\left[N^{(1...m)}_{(t-1)} = C \mid N^{(1...m)}_{(t)} = M\right]}{P\left[N^{(1...m)}_{(t-1)} = C\right]} * P\left[N^{(1...m)}_{(t)} = M\right] +$$

$$\frac{P\left[N^{(1...m)}_{(t-1)} = S \mid N^{(1...m)}_{(t)} = M\right]}{P\left[N^{(1...m)}_{(t-1)} = S\right]} * P\left[N^{(1...m)}_{(t)} = M\right]$$

$$= \frac{c^{1.m} * c^{1.m}}{1 - (\pi_1 + \pi_2 + \pi_3)} * \pi_2 + \frac{c^{1.m} * c^{1.m}}{1 - (\pi_0 + \pi_2 + \pi_3)} * \pi_2$$

$$x^{(1...m)} = \frac{P\left[N^{(1...m)}_{(t-1)} = C \mid N^{(1...m)}_{(t)} = F\right]}{P\left[N^{(1...m)}_{(t-1)} = C\right]} * P\left[N^{(1...m)}_{(t)} = F\right] +$$

$$\frac{P\left[N^{(1...m)}_{(t-1)} = S \mid N^{(1...m)}_{(t)} = F\right]}{P\left[N^{(1...m)}_{(t-1)} = s\right]} * P\left[N^{(1...m)}_{(t)} = F\right] +$$

$$\frac{P\left[N^{(1...m)}_{(t-1)} = M \mid N^{(1...m)}_{(t)} = F\right]}{P\left[N^{(1...m)}_{(t-1)} = M\right]} * P\left[N^{(1...m)}_{(t)} = F\right]$$

$$= \frac{d^{1...m} * d^{1...m}}{1 - (\pi_1 + \pi_2 + \pi_3)} * \pi_3 + \frac{d^{1...m} * d^{1...m}}{1 - (\pi_0 + \pi_2 + \pi_3)} * \pi_3 + \frac{d^{1...m} * d^{1...m}}{1 - (\pi_0 + \pi_1 + \pi_3)} * \pi_3$$

(11)

## 5. Model Parameters

This section describes how to obtain transition probabilities which are described as *a, b, c, d* and *e* above. To determine $P_{ij}$ (i,j ∈ {S} ), we need the appropriate input parameters from routing packets to determine its behavior and network resilience. Whenever node initiates the link, it will periodically broadcast a message to update its link information with its neighbors [20]. The message carries routing information such as number of packets, number of bytes sent and received, and its residual energy. Initially, every node has the same initial energy and may turn off packet forwarding functionality once its residual energy below a 1/η of its initial energy. The nodes tend to drop all the packets forwarded and become selfish at a time $T_{selfish}$ a s given below:

$$T_{selfish} = (1 - 1/\eta)\ \overline{L}$$

(12)

where η is the selfish threshold parameter and $\overline{L}$ is the average life time of a node derived from:

$$\overline{L} = \frac{\text{Remaining power}}{\text{Power consumption rate}}$$

(13)

Thus, the probability of selfish node a=1/$T_{selfish}$ and is given as:

$$a = \frac{\eta}{(\eta - 1)} * \frac{1}{\overline{L}}$$

(14)

The node is considered cooperative if 'a' is less than threshold and the average life time of the node will be longer. Similarly, the probability 'b' can be derived from monitoring neighboring nodes in terms of forwarding each other packets. Probability of 'b' can be derived as:

$$b = \frac{\text{Number of packets forwarded by the neighbors}}{\text{Number of packet received by the neighbors}}$$

(15)

The nodes may also receive more packets that it supposed to. This scenario considers a misbehavior activity by nodes. It is considered malicious nodes which goal is to depleting the resources of nodes forwarding packets. Probability 'c' can be derived from:

$$c = \frac{\text{Number of packet received by the neighbors}}{\text{Number of packets forwarded by the neighbors}} \quad (16)$$

The nodes consider malicious when 'c' is greater than threshold. Injecting an overwhelming amount of traffic also can easily cause network congestion and decrease the node lifetime. Thus, the node tend to loss it packets when the energy power is exhausted. Therefore, probability of loss 'd' is given as below:

$$d = \frac{b}{(\text{Num of packets forwarded} + \text{Num of packets receives})} \quad (17)$$

In the case when the node failed, it can be back in the network after the recovery time. The node recovery time can be derived using formula:

$$e = \frac{1}{\text{Average Recovery}} \quad (18)$$

## 6. Simulation and Analysis

In this paper we focus on the effect nodes correlation towards survivability. This section will verify the correctness and rationality of the equation discuss above. We use MATLABv7.10 to perform the simulation based on parameters discussed above. The survivability analysis in MANETS is considered from the following cases:

6.1 The effect on survivability from probability of cooperation (b)

As explain in section 3.0, node cooperation is represented by probability of forwarding $b$. The higher the $b$ implies that the node is cooperative. To observe the effect of probability of cooperation clearly, $c + d + e = 0$ is set so that $b$ varies only due to node selfishness attack. Figure 5 shows the analytic results of survivability under different nodes range 5, 15, 25, 50 nodes respectively. It is observed that cooperation level incline steady line with fewer nodes. This is due to the misbehavior node effect are less. Thus, the effect of node behavior is tractable with fewer nodes. This result proofs the concept from [7]. Cooperative nodes are affected by packet forwarding rate to obtain a higher survivability. Thus it is necessary to have a higher packet forwarding rate in order for network to survive. When drop packets are higher, the nodes become less cooperative and network survivability is impossible to achieve.

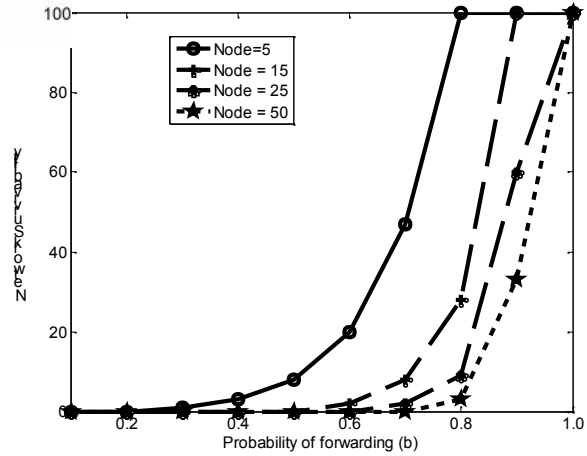

Fig.5: Effect on survivability of node cooperation based on probability of forwarding

6.2 The effect of survivability from the probability of selfishness (a)

In the same way, $c+d+e = 0$, and the result can be obtained from selfish nodes in Fig.6. Similar to that in Fig 5, the plot in Fig 6 shows that the survivability decreases as probability of dropping ($a$) increases. The survivability does not change significantly at the beginning especially if network scalability is less. In contrast, survivability for fewer nodes starts to decline faster compared to networks with large nodes. Network also becomes unstable when the dropping level reaches 0.4 to 0.5.

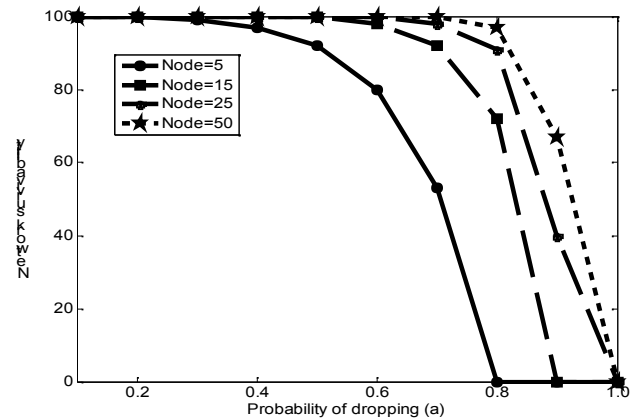

Fig 6.: Effect on survivability of node cooperation based on probability of dropping (a)

### 6.3 The effect on survivability from probability of malicious nodes (c).

To explain the effect of malicious node, a+d=0 is set to eliminate the impact of selfish and failure node. From Fig.7 the network survivability decreases very fast as probability of injection increase. It can be seen that network with more nodes could not sustain it survivability when network under attacks. When more packets were injected in the network, it will create a Denial of services (DoS) attack. DoS attack spread to other neighbors faster and will decrease network survivability faster too.

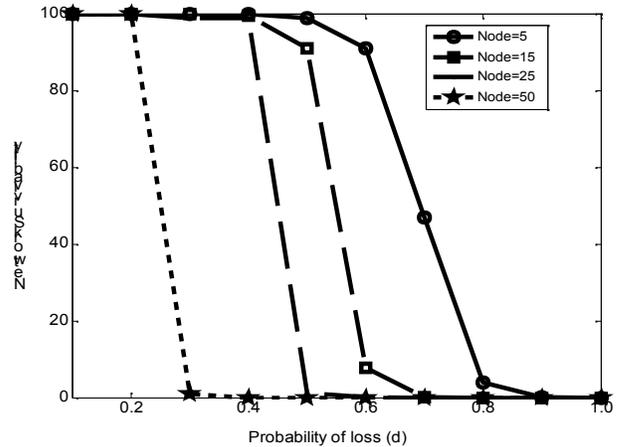

Fig 8. : Effect on survivability of node cooperation based on probability of loss (d)

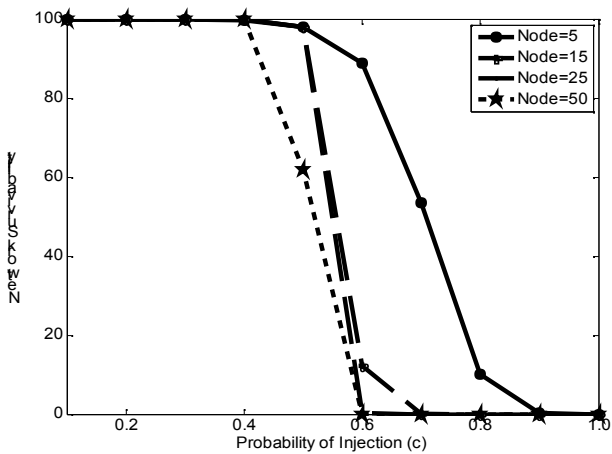

Fig 7. : Effect on survivability of node cooperation based on probability of injection (c)

### 6.4 The effect on survivability from probability of failure (d).

To explain the effect of node failures on network survivability, it is set that a+c=0 to eliminate the effect of misbehaving nodes in simulations. From Fig 8, survivability decreases very fast as probability of loss increase. For highly survival network, the effect of loss packet is more significant, e.g., the survivability drop to 0 when probability of packet loss (d) = 0.3. Compare to malicious nodes, failed nodes shown severe effect on network survivability. The severer impact of node failures is due to the fact that node failures are also isolated from the network, which reduces the density of active nodes [ ]. Therefore, the probability of network failure cannot be ignored especially for a large scale network.

## 5. Conclusions

In this paper, we focused on modeling and formulation correlated node behavior for dynamic topology network such as in MANETS. The nodes are classified into four types: cooperative, selfish, malicious and failed. Then, a correlated node behavior model was proposed by employing semi Markov process. The nodes were correlated according to their clusters describe as forwarding, dropping, injecting and loss. In our model, nodes changed their behavior according to correlated transition probability matrix and transition time distribution matrix. Further, the equilibrium states of correlated node being in each behavior state were obtained. The model showed accuracy in portraying the behavior of correlated node and thus can be used to study various performance metrics for network survivability and resilience. For future work, there is a need to solve the complexity of equation involving correlated transition probability matrix of $O(4^m)$ to simplify the computational processing and increase quality of services.

### Acknowledgments

A.H.Azni would like to thank Universiti Sains Islam Malaysia (USIM) and Ministry of Higher Education (MOHE) for financial support throughout her studies in Universiti Teknikal Melaka Malaysia (UTEM), Melaka, Malaysia.

# References

[1] F. Xing and W. Wang, "Understanding Dynamic Denial of Service Attack in Mobile Ad hoc Networks," in *IEEE Military communication conference (MILCOM)*, 2006, pp. 1-7.

[2] P. Rai, "A Review of ' MANET ' s Security Aspects and Challenges '," *System*, pp. 162-166, 2010.

[3] F. Xing and W. Wang, "Modeling and analysis of connectivity in mobile ad hoc networks with misbehaving nodes," in *IEEE International Conference on Communications, 2006*, 2006, vol. 4, no. c, pp. 1879–1884.

[4] J. P. G. Sterbenz et al., "Resilience and survivability in communication networks: Strategies, principles, and survey of disciplines," *Computer Networks*, vol. 54, no. 8, pp. 1245–1265, Jun. 2010.

[5] M. M. Sivajothi and E. Naganathan, "Analysis of Reference Point Group Mobility Model in Mobile Ad hoc Networks with an Ant Based Colony Protocol," *Proceedings of the International MultiConference of Engineers and Computer Scientists*, vol. 1, 2009.

[6] V. R. V. D. G. Mohankumar, "Feature Analysis for Intrusion Detection in Mobile Ad-hoc Networks," *IJCSNS*, vol. 10, no. 9, p. 215, 2010.

[7] F. Xing, "Modeling, Design, and Analysis on the Resilience of Large-scale Wireless Multi-hop Networks," University of North Carolina, 2009.

[8] Y. Liu, C. Comaniciu, and H. Man, "Modelling misbehaviour in ad hoc networks: a game theoretic approach for intrusion detection," *International Journal of Security and Networks*, vol. 1, no. 3/4, p. 243, 2006.

[9] T. Sundararajan and A. Shanmugam, "Modeling the Behavior of Selfish Forwarding Nodes to Stimulate Cooperation in MANET," *International Journal*, vol. 2, no. 2, pp. 147-160, Apr. 2010.

[10] S. Buchegger, "Coping with Misbehavior in Mobile Ad-hoc Networks," Citeseer, 2004.

[11] R. Mahajan, M. Rodrig, D. Wetherall, and J. Zahorjan, "Experiences applying game theory to system design," *Proceedings of the ACM SIGCOMM workshop on Practice and theory of incentives in networked systems - PINS '04*, p. 183, 2004.

[12] S. K. Hwang and D. S. Kim, "Markov model of link connectivity in mobile ad hoc networks," *Telecommunication Systems*, vol. 34, no. 1-2, pp. 51-58, Dec. 2006.

[13] P. Pileggi and P. Kritzinger, "Semi-Markov Process Formulation of IEEE 802 . 11 DCF Analytic Performance Models," *Computer*.

[14] M. Zhao and W. Wang, "A novel semi-markov smooth mobility model for mobile ad hoc networks," in *Proc. of IEEE GLOBECOM*, 2006, pp. 1–5.

[15] M. Fygenson, "SEMI-MARKOV PROCESSES," *Science*, vol. 32, pp. 151-160, 1989.

[16] V. G. Chaganti, L. W. Hanlen, and T. A. Lamahewa, "Semi-Markov modeling for body area networks," in *Communications (ICC), 2011 IEEE International Conference on*, 2011, pp. 1–5.

[17] D. N. Alparslan and K. Sohraby, "A Generalized Random Mobility Model for Wireless Ad Hoc Networks and Its Analysis: One-Dimensional Case," *IEEE/ACM Transactions on Networking*, vol. 15, no. 3, pp. 602-615, Jun. 2007.

[18] S. Wang and J. T. Park, "Modeling and analysis of multi-type failures in wireless body area networks with semi-Markov model," *Communications Letters, IEEE*, vol. 14, no. 1, pp. 6–8, Jan. 2010.

[19] Daniel Heyman and Mattlew Sobel., *Stochastic Models in Operations Research.* McGraw-Hill, 1982.

[20] K. Komathy and P. Narayanasamy, "A Probabilistic Behavioral Model for Selfish Neighbors in a Wireless Ad Hoc Network," *IJCSNS*, vol. 7, no. 7, p. 77, 2007.

**A.H. Azni** is a Ph.D student at Faculty of Information and Communication Technology, Universiti Teknikal Malaysia Melaka, with financial support from Ministry of Higher Education Malaysia and Universiti Sains Islam Malaysia. Azni received her M.sc in Digital Communication at Monash University, Clayton Australia in 2002 and B.sc in Computer Information System at Bradley University, Illinois, USA in 1998. She has work as a lecturer in Information and Security Assurance for many years, her interest domain is security, wireless and mobile network, survivability and resilience. Email: ahazni@usim.edu.my

**Rabiah Ahmad** is an Associate Professor at theFaculty of Information Technology and Communication, University Technical Malaysia Melaka (UTeM), Malaysia. She received her PhD in Information Studies (health informatics) from University of Sheffield, UK, and MSc. (information security) from Royal Holloway University of London, UK. Her research interests include healthcare system security and information security architecture. She delivered papers at various health informatics and information security conferences on national as well as international level. She has also published papers in accredited national/international journals. Besides that, she also serves as a reviewer for various conferences and journals. Email: rabiah@utem.edu.my

**Zul Azri Muhamad Noh** is a lecturer at Faculty of Information and Communcation Technology, Universiti Teknikal Malaysia Melaka, Malaysia. He received his D.Eng. (Computer Science and Engineering), M.Eng. (Computer Science and Engineering), and B.Eng. (Electrical and Computer Engineering) at Nagoya Institute of Technology, Nagoya Japan in 2005, 2007, and 2010, respectively. His current research interests include wireless LAN, packet scheduling, and multimedia QoS. Email: zulazri@utem.edu.my

**Abd Samad Hasan Basari** is a lecturer at Faculty of Information and Communcation Technology, Universiti Teknikal Malaysia Melaka, Malaysia. He received his B.Sc. (Mathematics) at National University of Malaysia in 1998, M.Sc in IT Education at Universiti Teknologi Malaysia in 2002 and Ph.D in Modeling at Universiti Teknikal Melaka Malysia in 2004y. His current research interests include Decision Support Technology, Modelling, Intelligent System Email: abdsamad@utem.edu.my

**Burairah Hussin** is an Associate Professor at the Faculty of Information Technology and Communication, University Technical Malaysia Melaka (UTeM), Malaysia. He received his PhD in Management Science from University of Salford, UK, and MSc. (Numerical Analysis and Programming) from University of Dundee, UK. His research interests include operational problem within industry, maintenance management, statistical analysis, creative media and software development. He has also published papers in accredited national/international journals. Besides that, he also serves as a reviewer for various conferences and journals. Email: burairah@utem.edu.my